# The magnetic ground state of a non-symmorphic square-net lattice, $TbAgSb_2$

Saswata Halder and Kalobaran Maiti*

*Department of Condensed Matter Physics and Materials Science, Tata Institute of Fundamental Research, Homi Bhabha Road, Colaba, Mumbai - 400 005, INDIA.*

*Corresponding author: kbmaiti@tifr.res.in

**Abstract.** We have investigated the magnetic properties of a non-symmorphic correlated material, $TbAgSb_2$. It consists of a quasi-two-dimensional structure of Sb atoms which hosts topologically non-trivial fermions. We prepared high quality single crystals of the material. The magnetic and transport properties exhibit antiferromagnetic transition at 11.3 K. A strong magneto-crystalline anisotropy is observed with the magnetic moments preferentially aligned within the ab-plane at low temperatures. A broad peak in the magnetic susceptibility along H||c direction suggests influence of crystal electric field in magnetism, which is also manifested in the specific heat data. The temperature dependent electrical resistivity at high field shows interesting signature of field induced magnetic structure reorientation / superzone formation. These results highlight the significance of Sb square nets in a non-symmorphic correlated system exhibiting complex magnetism and electronic properties.

## INTRODUCTION

The non-symmorphic square-net lattice $RAgSb_2$ (R = rare earth) hosts an interplay between localized rare-earth 4*f* spins and itinerant electrons, forming a heavy fermion state with a huge density of states close to the Fermi energy, $\varepsilon_F$ [1-3]. The $RAgSb_2$ series of materials crystallizes in a layered tetragonal structure with the rare-earth atoms residing close to the square-net Sb planes (Fig. 1). The crystal structure along the [001] direction follows a stacking order R-Sb1-Ag-Sb1-R sandwiched between two Sb2 square nets. Crystallographic non-symmorphic symmetry arising from the fractional displacement of R-atoms separated by a glide plane of the square net structure gives rise to protected band crossings near $\varepsilon_F$, which makes these materials a good platform to study the behavior of Dirac fermions in proximity to 4*f* moments [2,4]. Angle-resolved photoemission spectroscopy (ARPES) show distinct crossings of the Dirac-like linear bands near $\varepsilon_F$ in both $LaAgSb_2$ and $CeAgSb_2$ which are primarily constituted by Sb-5*p* orbitals from the square-net structure [4,5], which are robust even under strong spin-orbit coupling (SOC). Hard x-ray photoemission (HAXPES) investigation of an analogous AFM compound, $CeAgAs_2$, revealed interesting surface-bulk differences, a complex interplay of intra- and inter-layer covalency, and electron correlation in the electronic structure [6]. $SmAgSb_2$, another AFM material, shows well-resolved quantum oscillations that persist well into the paramagnetic domain [7-9]. This reveals the existence of an exotic symmetry protected topological band structure which is altered by the magnetic ground state [8,9]. Thus, the $RAgSb_2$ series of materials provides a fascinating playground to investigate the interplay between electron correlation from increased rare-earth 4*f* electron occupation and consequent decrease in 4*f*-hybridization in the presence of crystalline non-symmorphic symmetry.

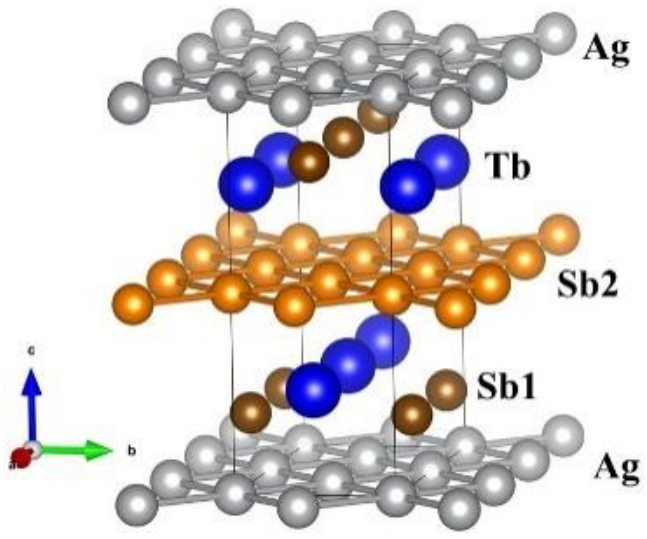

**Figure 1**. Crystal Structure of $TbAgSb_2$. Tb atoms are denoted by blue circles. Sb1 (brown) and Sb2 (orange) occupy the two non-equivalent Wyckoff positions. Ag atoms form the other dense square net arrangement.

In this work, we investigate the magnetic properties of $TbAgSb_2$ (TAS) which crystallize in the layered tetragonal $ZrCuSi_2$ structure with space group P4/*nmm* (#129) (Fig. 1a); Tb 4*f* band has significantly higher binding energy leading to a weaker 4*f*-hybridization relative to the Ce 4*f* band.

## CRYSTAL GROWTH AND CHARACTERIZATION

High-quality single crystals of TAS were prepared using the Sb-self flux method [1]. Tb metal chunks along with Ag and Sb metal shots were loaded inside a round bottom alumina crucible and sealed in a quartz ampoule under an inert environment. The ampoule was kept at 1050°C for 24 hours, followed by slow cooling (1°C/hour) down to 650°C at which the ampoule was centrifuged to remove the excess flux. The chemical composition of the compound was then characterized using energy-dispersive X-ray (EDX) spectroscopy. The magnetic properties were measured using a Quantum Design magnetic property measurement system (MPMS) under different temperature and magnetic field environment. The transport properties were measured using a Quantum Design physical property measurement system (PPMS) at different temperatures and magnetic fields.

## MAGNETIC AND TRANSPORT PROPERTIES

The magnetic dc susceptibility ($\chi$) of TAS as a function of temperature at H = 0.1 T along different crystallographic directions is shown in Fig. 2(a). The susceptibility data shows an AFM transition at $T_N$ = 11.3 K [1]. The magnetic properties are strongly anisotropic with the ab-plane being the easy plane of magnetic ordering. The $Tb^{3+}$ moments orient preferentially in this plane at low temperatures, with the crystallographic c-axis perpendicular to the basal plane being the hard direction for magnetic ordering. The inset in Fig. 2(a) shows the Curie-Weiss fit to the dc susceptibility data. The calculated effective moment of 9.37 $\mu_B$/Tb and 10.2 $\mu_B$/Tb which is close to the theoretically calculated moment ($\mu_{eff}$ ~9.72 $\mu_B$) for a free $Tb^{3+}$ ion. The non-linearity in the data below 100 K arise from the strong crystal electric field (CEF) effect. The negative sign of the Curie-Weiss temperature is consistent with the AFM exchange interactions in TAS. The different $\theta_p$ values along H||c ($\theta_p$ = -146.2 K) and $H \perp c$ ($\theta_p$ = -9.72 K) highlights strong anisotropy in the material. The magnetic susceptibility obtained at different magnetic fields is shown in Fig. 2(b) for H||a and H||c (inset). The AFM ordering is more prominent along the easy plane of magnetization. Although the AFM ordering is inconspicuous in the low field magnetization data along H||c, a broad peak in the magnetization curve centered at ~ 60 K is visible which is consistent with the strong anisotropic CEF.

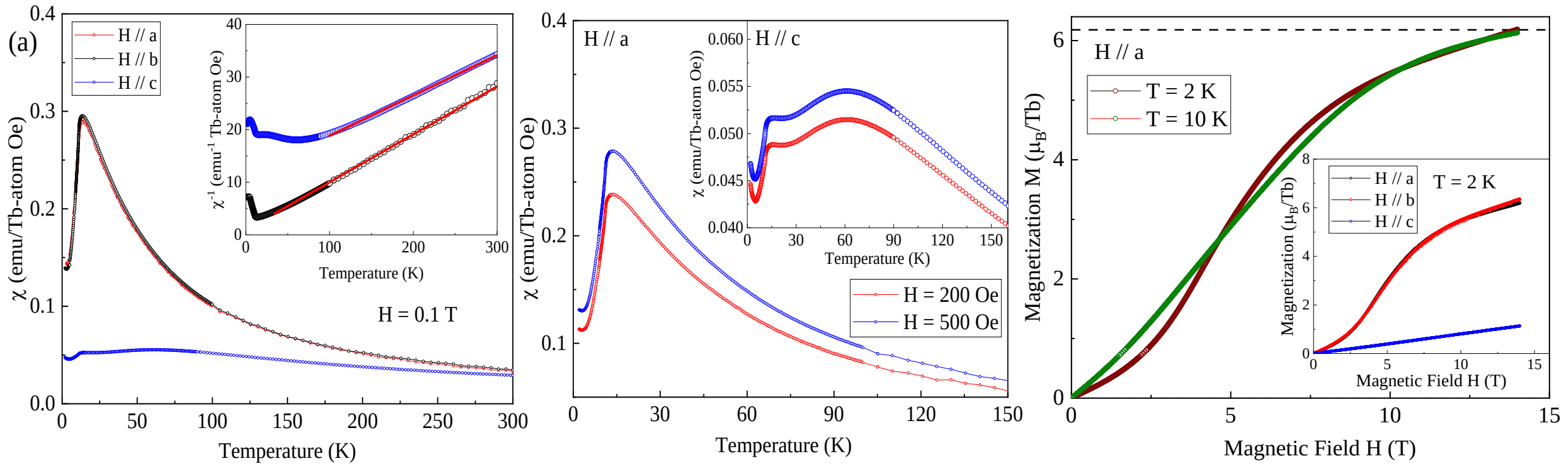

**Figure 2**. (a) Temperature dependent dc susceptibility along different crystallographic direction at H = 0.1 T. The inset shows the Curie-Weiss fit (b) dc magnetic susceptibility at different applied magnetic fields (c) The isothermal magnetization as a function of magnetic field for T = 2 K and 10 K respectively.

The isothermal magnetization as a function of the applied magnetic field for two different temperatures are shown in Fig. 2(c). For H||c (inset), the magnetization increases linearly with the applied magnetic field, reaching a

maximum saturation limit of 1.2 $\mu_B$/Tb at H = 14 T, consistent with strong magneto-crystalline anisotropy present in the material. Along the easy magnetization plane, a non-collinear response is observed with a broad meta-magnetic-like behavior at ~ 3 T (brown line). This meta-magnetic-like behavior smoothens out with increasing temperature (green line). The saturation moment for the M-H isotherm at 2 K remains well suppressed at ~6.2 $\mu_B$/Tb much less than the saturation moment for a free $Tb^{3+}$ ion. This points towards the possibilities of further meta-magnetic transitions at higher fields.

The temperature-dependent heat capacity, $C_p$ is shown in Fig. 3(a). The experimental data show the signature of AFM transition at ~11.3 K which is consistent with the magnetization measurements. However, the spectra saturate well below 3NR due to the strong CEF effects present in TAS. The variation of $C_p$/T with temperature (Fig. 3a inset) shows a broad hump in low temperature region close to $T_N$. In order to extract the magnon contribution ($C_{mag}$), the phonon background is subtracted using the specific heat data of non-magnetic $LaAgSb_2$ (Fig. 3b). The obtained $C_{mag}$ is shown in the inset of Fig. 3(b). The magnon peak tends to remain quasi-static along the temperature axis showing the presence of strong exchange interactions in TAS. The thermal variation of the electrical resistivity of TAS for different magnetic fields is shown in Fig. 3(c). The resistivity data show the signature AFM ordering along with a linear increase at higher temperatures. The residual resistivity ratio (RRR = ρ(300 K)/ρ(2 K)) is 45, which indicates high purity and low dislocation densities in the measured samples. Below $T_N$, a clear loss of spin-disorder scattering corresponding to the AFM transition is visible. At higher magnetic fields (> 5 T), the resistivity data shows a change in behavior corresponding to the change in the spin arrangement in the AFM lattice.

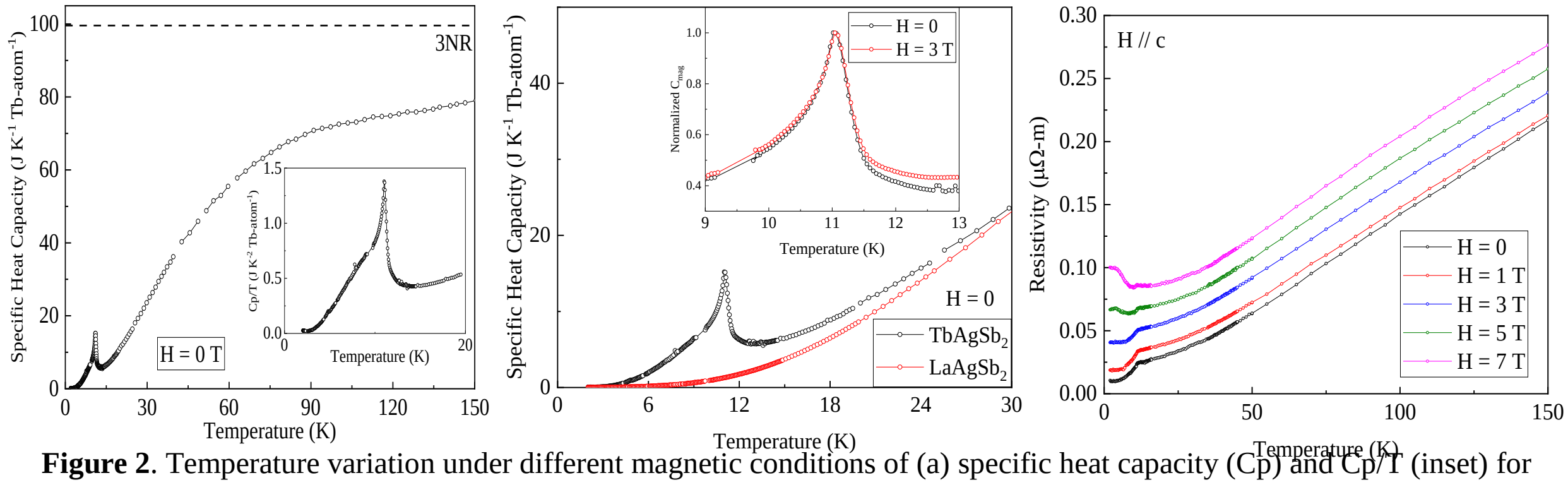

**Figure 2**. Temperature variation under different magnetic conditions of (a) specific heat capacity (Cp) and Cp/T (inset) for $TbAgSb_2$ (black). (b) $C_p$ for $TbAgSb_2$ and $LaAgSb_2$. The inset shows the variation for $C_{mag}$ (c) electrical resistivity under different applied magnetic fields.

In summary, we investigated the magnetic and transport properties of TAS under varying temperatures and magnetic field environments. TAS is a layered quasi-2D material that consists of two inequivalent Sb sites; one of which hosts a 2D square net structure. The magnetic properties show the signature of an AFM ground state with strong magneto-crystalline anisotropy. The basal ab-plane emerges as the easy plane of magnetization with moments confined to this plane. The magnetization isotherms show a weak metamagnetic-like behavior, with additional high-field metamagnetic transitions in the high magnetic field regime probable. The heat capacity and electrical resistivity data also show the AFM phase boundary. The transport behavior as function of magnetic field appear unusual within the antiferromagnetic regime suggesting field induced super-zone formation in this system.